\documentclass[aps, prb, reprint, twocolumn, superscriptaddress]{revtex4-1}
\usepackage{graphicx} 
\usepackage{braket}
\usepackage{color}
\usepackage{times}
\usepackage{textcomp} 
\usepackage{amsmath}

\begin{document}

\title{A quantum fiber-pigtail}

\author{Davide Cadeddu}
\author{Jean Teissier}
\author{Floris Braakman}
\affiliation{Department of Physics, University of Basel, Klingelbergstrasse 82, CH-4056 Basel, Switzerland}
\author{Niels Gregersen}
\affiliation{DTU Fotonik, Department of Photonics Engineering, Technical University of Denmark, Building 343, DK-2800
Kongens Lyngby, Denmark}
\author{Petr Stepanov}
\author{Jean-Michel G\'{e}rard}
\author{Julien Claudon}
\affiliation{Université Grenoble Alpes, F-38100 Grenoble, France}
\affiliation{CEA, INAC-SP2M, 17 rue des Martyrs, F-38054 Grenoble, France}
\author{Richard J. Warburton}
\author{Martino Poggio}
\author{Mathieu Munsch}
\affiliation{Department of Physics, University of Basel, Klingelbergstrasse 82, CH-4056 Basel, Switzerland}

\date{ \today}

\begin{abstract}
We present the experimental realization of a quantum fiber-pigtail. The device consists of a semiconductor quantum-dot embedded into a conical photonic wire that is directly connected to the core of a fiber-pigtail. We demonstrate a photon collection efficiency at the output of the fiber of $5.8\,\%$ and suggest realistic improvements for the implementation of a useful device in the context of quantum information. We finally discuss potential applications in scanning probe microscopy.
\end{abstract}

\maketitle

Semiconductor quantum-dots (QDs) are attractive single photon sources. They are robust, compact and provide on-demand single photons at rates in the GHz range~\cite{Michler2000,Santori2001,Moreau2001}. Their potential in the context of quantum optics however relies on the fulfillment of several demanding criteria~\cite{Sangouard2012}: high efficiency, high photon purity and simple operation. Recent progress has nevertheless brought QDs close to such applications. 
Single-photon operation has been obtained in a compact, table-top Stirling machine~\cite{Schlehahn2015}, offering a low-cost and user-friendly solution. 
Thanks to the increasing quality of the epitaxial material, spectrally pure emission has been demonstrated~\cite{Kuhlmann2013a}. 
The last challenge that needs to be addressed is to efficiently couple the emitted light into a single mode fiber. 
Large progress in this direction has been made with the integration of QDs into micro and nano-scale photonic structures, such as cavities and waveguides, which allow the control of spontaneous emission~\cite{Gerard1998,Bleuse2011,LundHansen2008}.
In the last few years, important efforts to position the QD in an optimal way~\cite{Dousse2008, Reimer2012} and to minimize the diffraction of light at the output of photonic nanowires~\cite{Claudon2013} have pushed the collection efficiencies to values $ \gtrsim 75 \%$ while maintaining a Gaussian spatial profile~\cite{Gazzano2013, Munsch2013}. 
These impressive results require however the use of objective lenses with large numerical apertures. In parallel, different strategies to couple the emitted light directly into a single mode fiber have emerged~\cite{Xu2007c,Haupt2010,Tiecke2015}.

\begin{figure}[b!]
\includegraphics[width=0.48\textwidth]{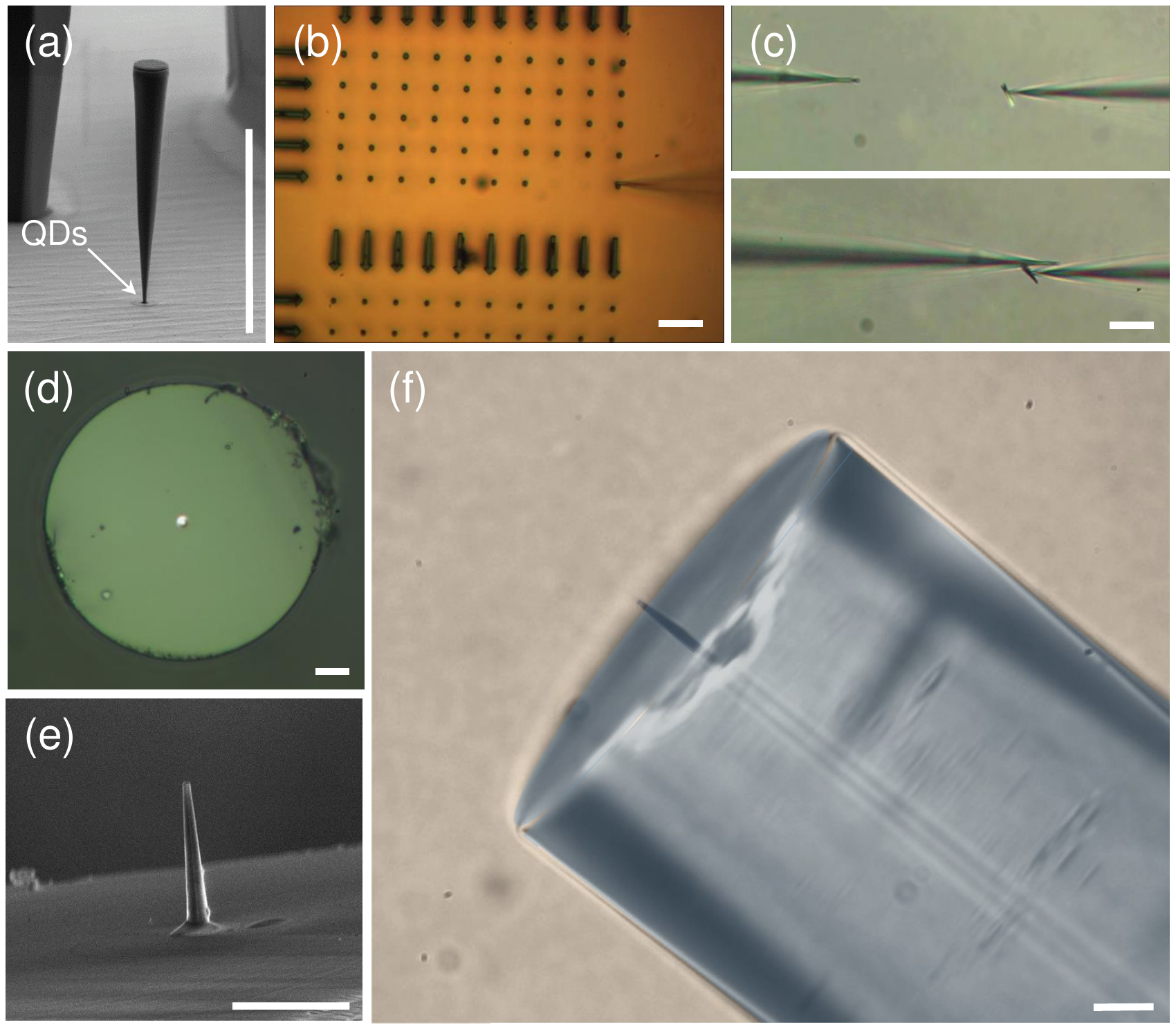}
\caption{{\bf Fabrication procedure}. a) SEM picture of the photonic trumpet after etching and removal of the Ni mask. b) Removing a single PW from its original substrate with glass micro-manipulator (right side). c) Orientation of the PW for subsequent gluing onto the fiber. d) Top view of the bare fiber with a drop of UV glue (bright spot) deposited at its center. e) SEM picture of the fiber-wire connection. The drop of glue can be seen at the base of the PW. f) Side view of the final device (optical microscope image). The white scale bars represent $10$ \textmu m.}
\label{Figure1}
\end{figure}

We report here the direct coupling of QD single photons to an optical fiber with a new approach. Our device, the quantum fiber-pigtail, consists of a QD embedded in a tapered photonic wire (PW), a photonic ``trumpet", that is directly attached to the cleaved end of a single mode fiber. Thanks to adiabatic expansion of the guided mode confined in the PW, we achieve an external collection efficiency of $5.8 \%$ at the output of the fiber-pigtail. The result represents a proof-of-principle for an easy-to-operate single photon source. We discuss realistic improvements and show that an efficiency exceeding $70 \%$ is within reach with current technology. Furthermore, easily addressable QDs at the end of a nanometer-scale tip have obvious potential as scanning probes. Possible applications include single photon near-field microscopy~\cite{Cuche2009}, deterministic quantum plasmonics~\cite{Cuche2010} or electric field sensing~\cite{Vamivakas2011}.

Our quantum emitter consists of a self-assembled InAs quantum dot (QD) grown by molecular beam epitaxy. It is embedded $110$ nm away from the sharp end of a $12$ \textmu m-long conical PW standing on a gold substrate~\cite{Munsch2013}, see Fig. \ref{Figure1}a.  
At the QD position, the small diameter enhances the coupling of the emitter to the guided modes while screening its coupling to lateral free-space modes~\cite{Bleuse2011}. As the top facet is approached, the progressive increase in diameter allows for an adiabatic expansion of the guided modes and eventually leads to a better mode matching with the fiber.
These conical wires are obtained through a top-down approach. Using e-beam lithography, we define a Ni hard mask consisting of arrays of disks with variable diameters. This is followed by a deep plasma etch conducted in a reactive ion-etching chamber. Finally, the remaining Ni mask is removed in a diluted nitric acid solution. A Si$_3$N$_4$ anti-reflection coating maximizes the transmission through the top facet. The resulting structures are shown on Figs. \ref{Figure1}a and \ref{Figure1}b.
 
\begin{figure}[t]
\includegraphics[width=0.48\textwidth]{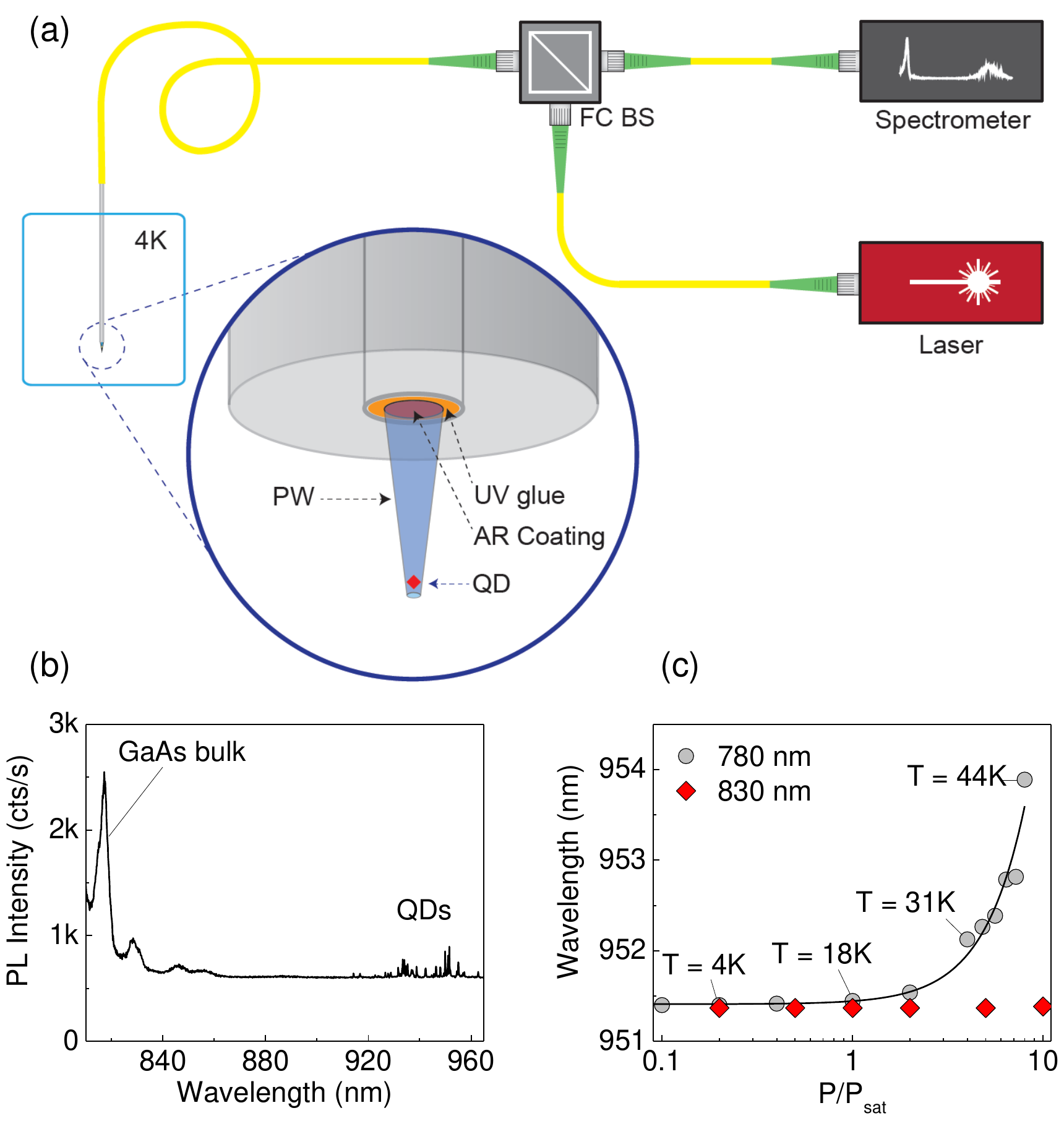}
\caption{{\bf General properties}. a)~The quantum fiber-pigtail is cooled down to 4 K in liquid He. The QDs are excited with a CW laser diode ($\lambda = 830$ nm / optionally $780$ nm, see text). The photoluminescence (PL) is analyzed with a spectrometer and a CCD camera (not shown). b)~Typical PL spectrum for excitation in the GaAs bulk. c)~Power-induced heating of the PW for excitation in the GaAs bulk (gray circles). Excitation in the wetting layer (red diamonds) results in negligible heating. The abscissa has been normalized to the saturation power $P_{\rm sat}$. The solid line is a quadratic fit to the data.}
\label{Figure2}
\end{figure}

In order to realize a direct coupling between the photonic trumpet and a single mode fiber, we pick up an individual wire and glue it to the core of a standard fiber ($d_{\rm core}\,=\,4.4$~\textmu m, $n_{\rm core}\,=\,1.4563$ and $n_{\rm cladding}\,=\,1.4513$), as illustrated in Fig.~\ref{Figure2}a. Initially, we fabricate micro-manipulators by tapering a glass needle down to a few micron thickness. The micro-manipulator is then used to pick up one wire at a time thanks to a combination of electrostatic and Van der Waals forces (see Fig.~\ref{Figure1}b).
Observations with a scanning electron microscope (SEM) indicate that the cleaving point lies at the interface between the PW and the substrate, within an estimated error of $\pm 10$ nm (the instrument's resolution).
With a second micromanipulator we re-orient the PW into the appropriate direction (top facet facing downwards), Fig.~\ref{Figure1}c. We then approach the cleaved facet of a fiber-pigtail, and deposit a drop of low fluorescence UV glue onto its core, Fig. \ref{Figure1}d. Finally, we bring the wire and the fiber into contact with an alignment precision on the order of 1 \textmu m and illuminate with UV light to harden the glue. The resulting structure is shown in Figs. \ref{Figure1}e and \ref{Figure1}f. It is robust and resistant to cycling to cryogenic temperatures\footnote{We performed up to 3 cycles with the same device without loss of signal}, two important points for future applications.

The device is tested at cryogenic temperature by plunging it directly into liquid He. The QDs are excited non-resonantly with a CW laser diode and the photoluminescence is analyzed with a spectrometer and a high efficiency Si-based CCD camera ($\eta_{det} = 27 \%$ at $\lambda = 950$ nm), see Fig.~\ref{Figure2}a. 
A typical spectrum is shown in Fig.~\ref{Figure2}b for an excitation in the bulk ($\lambda_{\rm laser} = 780$ nm). We identify a peak associated to the GaAs nanowire and a series of sharp lines corresponding to several QDs.
Importantly, we observe significant heating as we increase the non-resonant power. This is evidenced as a quadratic shift of the QD energies in Fig. \ref{Figure2}c, and indicates a poor heat dissipation in the device, despite the surrounding liquid He.
A simple way to avoid this problem is to create electron-hole pairs directly in the InGaAs wetting layer connecting the QDs ($\lambda_{\rm laser} = 830$ nm). In this case, we minimize the amount of absorbed light and observe no heating effect over the range of useful excitation powers (see Fig.~\ref{Figure2}c). This second scenario was used for all the results presented below.

\begin{figure}[tb]
\includegraphics[width=0.48\textwidth]{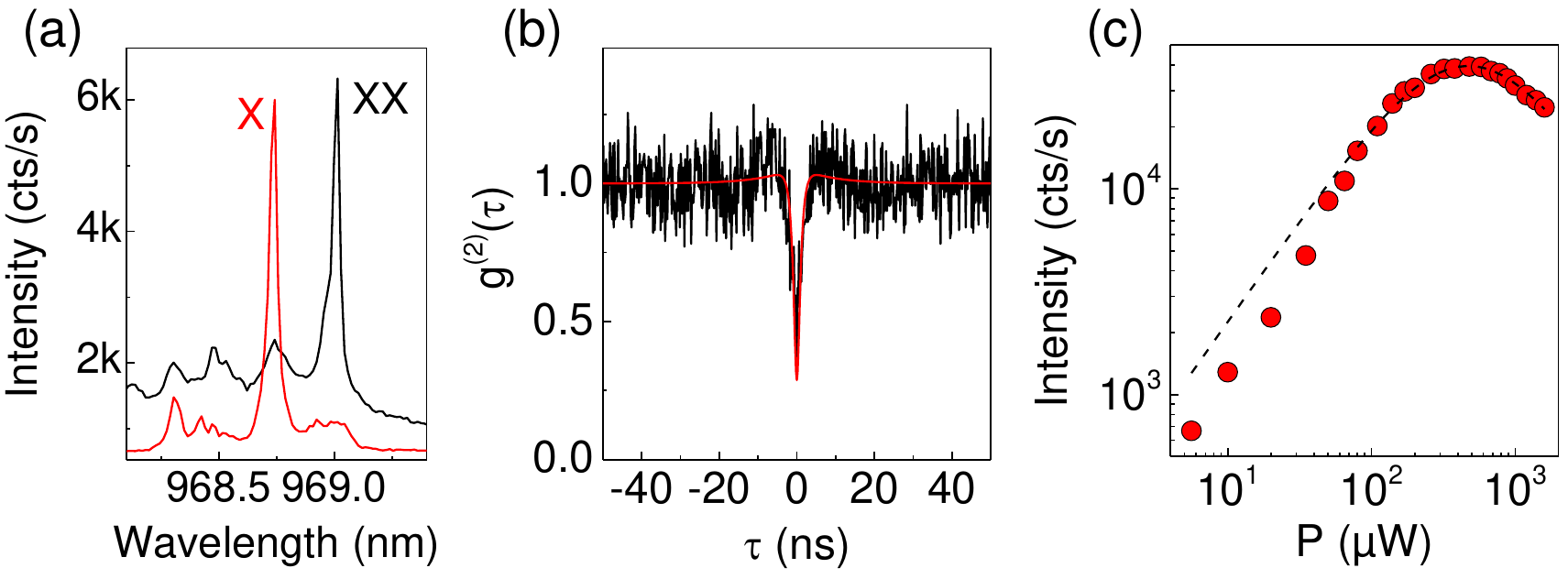}
\caption{{\bf Characterization of a single QD}. a)~QD1 spectra at low ($P=125$~\textmu W, in red) and high ($P=1.2$~mW, in black) powers. X and XX correspond to the exciton and the biexciton.  b)~Auto-correlation signal of X ($P = 64$ \textmu W). The data are normalized to the average coincidence counts per time bin. The dip at zero delay is the signature of a single quantum emitter. The red solid line is a fit to the 2-level atom result including the detector time-jitter ($\sim 400$ ps). c)~Power dependence of the exciton PL intensity. The dashed line is a fit using Eq. 1 from the main text.}
\label{Figure3}
\end{figure}

We focus on a PW featuring a diameter of $440$ nm at the QD's position and $1.8$ \textmu m at the top facet. The wire sustains the guided modes associated with the first six orders and contains approximately $50$ QDs distributed over a spectral bandwith of about $50$~nm. In the low energy tail, we identify in particular a bright complex labeled QD1, which consists of an excitonic transition X (possibly a charged exciton) and a red-shifted biexciton XX at high power (see Fig. \ref{Figure3}a).

To confirm the quantum nature of the emitted signal, we perform an auto-correlation measurement on the exciton using a standard Hanbury-Brown Twiss experiment with a set of two avalanche photo-diodes. The result, shown in Fig. \ref{Figure3}b, reveals a pronounced dip at zero delay, characteristic of anti-bunched emission. The data are very well reproduced by the auto-correlation function of a perfect 2-level emitter with a lifetime of $1/\gamma_X = 1.2$ ns convoluted with the detector's response (Gaussian with full-width-half-maximum of $\sim 400$ ps).

To evaluate the extraction efficiency $\epsilon$ of our quantum fiber-pigtail, we record the PL intensity as we increase the excitation power. As we saturate the X transition, we observe a maximum count rate of $\sim~40$~kcts/s, Fig. \ref{Figure3}c. The data is fitted using a simple three-level model that takes the biexciton into account. Denoting $\gamma_{\rm X}$ and $\gamma_{\rm XX}$ the decay rates for the exciton and the biexciton, the detected PL intensity is given by~\cite{Munsch2009}
\begin{equation}
I_{\rm X, det} (P) = \frac{I_{\rm sat}}{1+\frac{\alpha P}{\gamma_{\rm X}} +\frac{\gamma_{\rm XX}}{\alpha P}}
\end{equation}
with $I_{\rm sat}$ the intensity at saturation, and $\alpha$ a coefficient which translates the measured excitation power into an effective pumping rate. Using $\gamma_{\rm X} = 0.84$ GHz from the auto-correlation measurement and $\gamma_{\rm XX}\sim  2 \gamma_{\rm X}$, we obtain very good agreement with our experimental data for $I_{\rm sat}~=~149$~kcts/s and $\alpha~=~2.5~\times~10^{-3}$~GHz/\textmu W. 
The relationship between the detected flux and the emission rate simply reads
\begin{equation}
I_{\rm sat} = \epsilon \, \eta_t \,\eta_{\rm det} \, \gamma_{\rm X},
\end{equation}
where $\eta_t$ corresponds to the overall transmission between the fiber-pigtail and the detector. Using a reference tunable laser diode set at $970$ nm, we find $\eta_t = 6 \% \pm 3.5 \% $, which yields a collection efficiency $\epsilon = 5.8 \% \,\pm\, 3.3 \%$. The given value includes all losses, for instance the finite coupling of the QD to the waveguide-modes propagating in the upward direction, the imperfect wire-to-fiber mode matching and the transmission losses.
This result can still be improved, but we stress that it is already more than one order of magnitude superior to the value one would obtain from QDs in the bulk. It therefore constitutes a proof-of-principle for our approach of integrating a quantum light source to a standard optical fiber.

We estimate the quality of the fabrication by evaluating the expected maximum efficiency at the output of the fiber-pigtail. Using Lumerical FDTD Solutions, we calculate the fraction of power radiated through the top facet of the PW by an embedded dipole point-source. The result is multiplied by the overlap between the computed profile of the electromagnetic field (at the top facet) and the mode of the fiber. 
For a PW with the above-mentioned dimensions and a QD on axis, we find $\epsilon = 9.2 \,\binom{+0.4}{-0.7} \%$, where the error bar comes from the $\pm 10$~nm uncertainty on the distance between the QD and the cleaved facet. 
Our experimental result is in agreement with the theoretical prediction. This calculation shows that we have met the main challenges, namely positioning the QD close to the axis and aligning the center of the PW with the core of the fiber. 

We believe that the proposed device has the potential to relax some of the constraints associated with the use of single QDs in quantum optics experiment. Integrating it into a compact closed cycle cryostat would result in a quantum light source taking the form of a "black-box", with single photons directly available at the output of a standard single mode fiber. To be useful in a quantum optics context, a single photon source should be spectrally pure (usually a challenge for solid-state emitters) and present high collection efficiencies, typically higher than $\simeq 70 \%$ for a quantum repeater protocol based on single photons~\cite{Sangouard2012}. Such a quantum pigtail is not out of reach. 
Recent experiments indicate that the coherence of photons emitted by QDs embedded in such photonic wires does not suffer from the presence of nearby etched surfaces \footnote{\textit{Manuscript in preparation}}. Though these experiments were not carried out on the highest quality material (bulk QD linewidth $\sim 7$ times the Fourier transform limit), the results are very encouraging. 
We shall now show that high efficiencies are reachable with realistic improvements. 

To optimize $\epsilon$, a natural strategy is to operate the tapered PW in the single mode regime. This choice simultaneously ensures high emission probabilities into the fundamental guided mode (HE$_{\rm 11}$) and optimum mode matching between the PW and the fiber. In an PW with a diameter $0.22 < D/\lambda < 0.31$, more than $95 \%$ of the QD spontaneous emission is funneled into HE$_{\rm 11}$~\cite{Bleuse2011}. For such a symmetric waveguide, the emission is evenly distributed between the upward and downward directions. The fraction of photons emitted in the guided mode \emph{propagating towards the fiber} can be increased up to $\beta_{\rm up} \sim 72 \%$ by exploiting the reflectivity associated with a simple cleaved facet~\cite{Friedler2009}. This value can be further enhanced to $\beta_{\rm up} \sim 92 \%$ by depositing a metallic post mirror on the facet~\cite{Friedler2009}. In both cases, the QD should be located at an anti-node of the electric field to benefit from constructive interference~\cite{Friedler2009}. To optimize the mode matching between the PW and the fiber, the diameter of the PW's top-facet should be adjusted. We find a mode overlap $\mathcal{O}> 86\%$ when the top-facet diameter exceeds $7$ \textmu m. Importantly adiabatic conditions have to be maintained along the taper to minimize coupling to higher order modes propagating in the PW.
As an example, we consider a $78$~\textmu m long single mode PW with a tapering angle of $5^{\circ}$ ensuring a transmission of the fundamental mode $\mathcal{T}_{\rm HE_{11}} = 90 \%$. The top facet has a diameter of $7$~\textmu m, and the cleaved apex is covered with a silver mirror~\cite{Friedler2009}. 
With the active layer located $110$~nm above the mirror, we obtain a total efficiency $\epsilon_{\rm th} > 71 \%$. Remarkably, the simple cleaved facet with no additional mirror already results in an efficiency of $\epsilon_{\rm th} > 54 \%$. To conclude, we stress that one may also tailor the properties of the fiber and release some constraints on the PW geometry.

Our device also presents attractive features as a surface scanning probe. We point out two examples of possible applications along this line.

\begin{figure}[tb]
\includegraphics[width=0.4\textwidth]{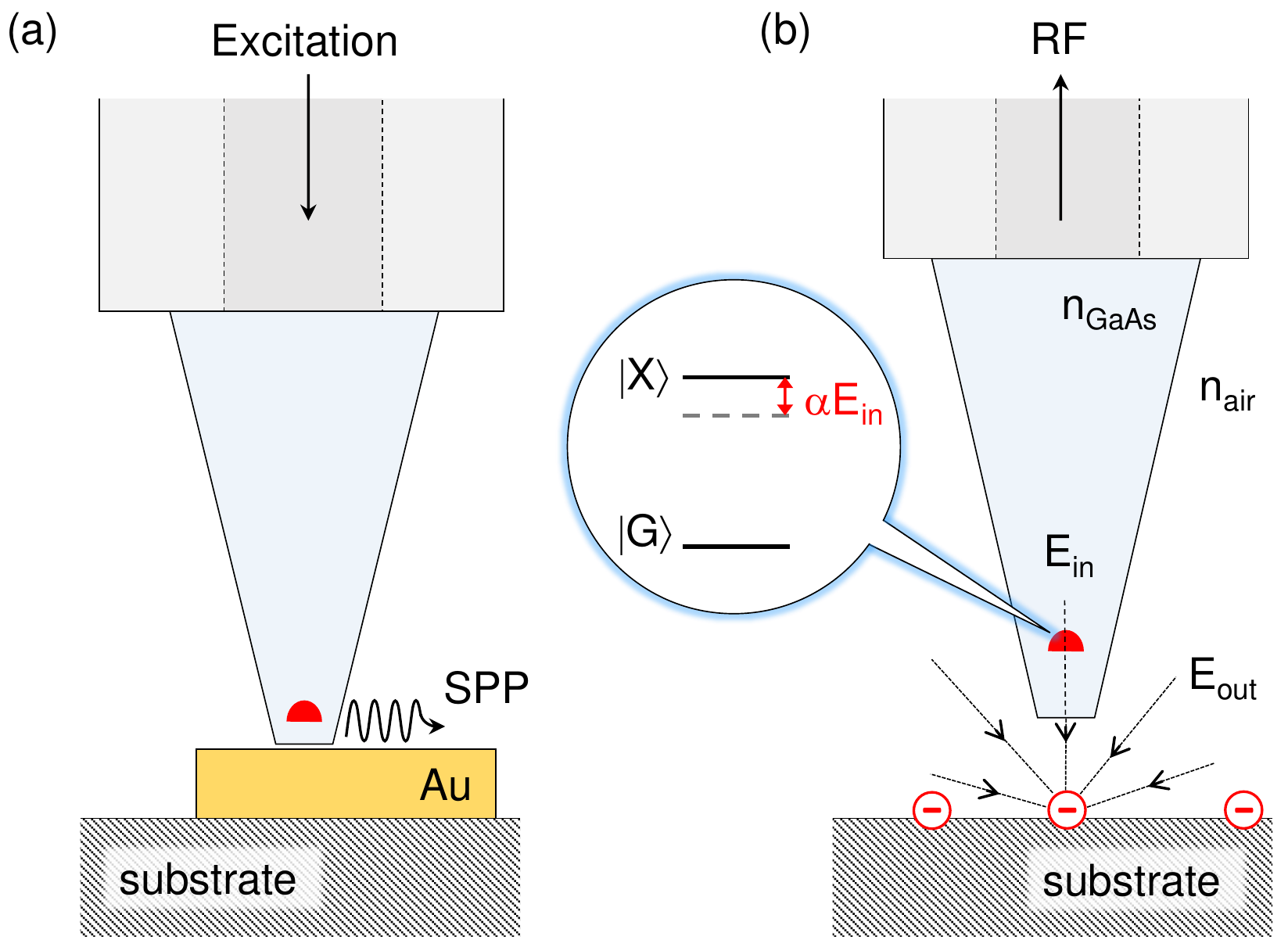}

\vspace{0.5cm}

\begin{tabular*}{\linewidth}{@{\extracolsep{\fill}}ccc}
\hline
\hline
& a) & b)  \\
\hline
 distance to & $ d \sim 30$ nm   & $d< 4$\textmu m$^{*}$\\
 substrate &   &  \\
 \hline
 figure of & $\beta_{\rm pl}$-factor & sensitivity $\eta$\\
merit & $ \sim 50\,\%$  & $20$ (V/m)/$\sqrt{\rm Hz}$\\
\hline
\hline
\multicolumn{3}{r}{\footnotesize{* sensitivity to a single electron charge for 1s integration time}}

\end{tabular*}
\caption{{\bf Quantum-dot based scanning probe microscope}. 
a) Deterministic quantum plasmonics. A QD is brought into close proximity with a metallic antenna. The resulting change in the local density of states leads to efficient plasmon launching. The numbers in the table correspond to the experimental results reported in \cite{Andersen2011}. $\beta_{\rm pl}$ corresponds to the probability of exciting a plasmon for one excitation of the QD. b) Electric field sensing. An electric field produces a Stark shift of the QD excitonic level which is detected via the resonance fluorescence intensity.
}
\label{Figure4}
\end{figure}

The first one concerns quantum plasmonics. Surface plasmon polaritons (SPPs) represent a possible way of building integrated quantum optics circuits at the nanoscale~\cite{Tame2013}. Our device could be used to transfer quantum information from the QD to propagative SPPs simply by bringing the sharp tip of our photonic wire into close proximity with a metallic nanostructure ($d < \lambda/2$), Fig. \ref{Figure4}a. 
Compared to previous work~\cite{Akimov2007,Bracher2014}, this solution presents more flexibility and would allow scanning of the sample surface or bringing the quantum emitter to a specific location, while controlling its exact distance to a given metallic antenna.
The possibility of positioning our probe at will presents a significant advantage to explore the effect of the near-field environment on the emission properties of a QD. 
In particular, recent experiments have shown that control over the distance (as well as over the QD orientation) could lead to launching plasmons with probabilities approaching $50\,\%$~\cite{Andersen2011}.
The present device is thus likely to find applications in the field of quantum plasmonics, in particular when more complex plasmonic circuits come to the fore~\cite{Fang2010,Wei2011a}.

A second application concerns using the QD at the apex of the wire as an ultra-sensitive electric field sensor, Fig. \ref{Figure4}b.
The idea is that an electric field $E_{\rm in}$ results in a Stark shift that can be detected through a change in intensity of QD resonance fluorescence. The noise spectrum of the QD resonance fluorescence signal reveals the noise spectrum of the local electric field~\cite{Kuhlmann2013a}.
In the linear regime and for a vertical electric field, the Stark shift is given by $\Delta \nu = \alpha^{-1} E_{\rm in} /h$, where $\alpha \sim 0.3$~\textmu eV/(kV/m)~\cite{Houel2012,Vamivakas2011} and $h$ is the Planck constant. Assuming a detection floor limited by shot noise, we derive the sensitivity of the QD sensor
\begin{equation}
\eta = \frac{\alpha^{-1}}{\partial \dot{n}(\nu)/\partial \Delta\nu} \sqrt{2\dot{n}(\nu)},
\end{equation}
where $\dot{n}(\nu) = \frac{\dot{n}_0}{1 +( 2\nu /\Gamma)^{2}}$ is the resonance fluorescence count rate. Assuming a count rate at resonance $\dot{n}_0 =1$ MHz, a linewidth of $\Gamma = 5$ \textmu eV, and taking into account the reduction in electric field amplitude due to its penetration into a high refractive index material ($n_{\rm GaAs}=3.5$)
\begin{equation}
E_{\rm in} \propto \frac{1}{n^2} E_{\rm out},
\end{equation}
we obtain sensitivities to an external electric field  $E_{\rm out}$ as high as  $\eta \sim 20$ (V/m)/$\sqrt{\rm Hz}$. This corresponds to detecting a single charge at a distance of $4$ \textmu m within a 1~s integration time, or alternatively to detecting a single charge at a distance of $150$~nm in just 1~ms. Importantly, we point out the very large sensor bandwidth, which is ultimately limited by the spontaneous emission rate and can thus approach the GHz range in the case of standard InGaAs QDs.

QD electric field sensors display sensitivities which compete with those of a single-electron transistor, while minimally affecting the system being detected \cite{Vamivakas2011}. The scanning capability offered by our quantum pigtail makes it an appealing alternative to existing devices. We finally note that the quantum pigtail could be straightforwardly attached to a tuning fork force sensor~\cite{Karrai1995} facilitating atomic force microscopy with the tapered photonic wire as tip, and that the translation of this concept to NV centers in diamond would enable the creation of a scanning magnetic field sensor~\cite{Degen2008}.

In conclusion, we present a device concept for an alignment-free, easy-to-operate source of single photons. A semiconductor photonic trumpet containing quantum dots is attached to the facet of a single-mode optical fiber. As a figure of merit we have evaluated the extraction efficiency at the output of the fiber-pigtail ($5.8\,\%$), and demonstrated good agreement with numerical calculations. We have shown that realistic adjustment could lead to values $> 70\,\%$. Importantly the fabricated device is robust and operates well in a cryogenic environment. Finally, we have underlined its potential in the context of deterministic quantum plasmonics and electric field sensing.

\section*{Funding Information}
This work was supported by an ERC Grant (NWscan, Grant No. 334767), the Swiss Nanoscience Institute (Project P1207), and the NCCR QSIT. JC and JMG acknowledge the support of the European Union Seventh Framework Program 209 (FP7/2007-2013) under Grant Agreement No. 601126 210 (HANAS) and the European Metrology Research Program (EMRP) [project SIQUTE (contract EXL02)]. NG acknowledges support from the Danish Research Council for Technology and Production (project LOQIT, grant no. 4005-00370B). The photonic trumpet fabrication was carried out in the 'Plateforme technologique amont' and CEA LETI MINATEC/DOPT clean rooms.

\section*{Acknowledgments}

We thank Andrea Mehlin for technical assistance, Nicolas Sangouard and Patrick Maletinsky for discussions.

\section*{Supplemental Documents}
A video illustrating the fabrication process is included.


\bibliography{munsch_refs_april_2015}

\begin{thebibliography}{10}
\expandafter\ifx\csname url\endcsname\relax
  \def\url#1{\texttt{#1}}\fi
\expandafter\ifx\csname urlprefix\endcsname\relax\def\urlprefix{URL }\fi
\providecommand{\bibinfo}[2]{#2}
\providecommand{\eprint}[2][]{\url{#2}}

\bibitem{Michler2000}
\bibinfo{author}{Michler, P.} \emph{et~al.}
\newblock \bibinfo{title}{A quantum dot single-photon turnstile device}.
\newblock \emph{\bibinfo{journal}{Science}} \textbf{\bibinfo{volume}{290}},
  \bibinfo{pages}{2282} (\bibinfo{year}{2000}).

\bibitem{Santori2001}
\bibinfo{author}{Santori, C.}, \bibinfo{author}{Pelton, M.},
  \bibinfo{author}{Solomon, G.}, \bibinfo{author}{Dale, Y.} \&
  \bibinfo{author}{Yamamoto, Y.}
\newblock \bibinfo{title}{Triggered single photons from a quantum dot}.
\newblock \emph{\bibinfo{journal}{Phys. Rev. Lett.}}
  \textbf{\bibinfo{volume}{86}}, \bibinfo{pages}{1502} (\bibinfo{year}{2001}).

\bibitem{Moreau2001}
\bibinfo{author}{Moreau, E.} \emph{et~al.}
\newblock \bibinfo{title}{Single-mode solid-state single photon source based on
  isolated quantum dots in pillar microcavities}.
\newblock \emph{\bibinfo{journal}{Applied Physics Letters}}
  \textbf{\bibinfo{volume}{79}}, \bibinfo{pages}{2865--2867}
  (\bibinfo{year}{2001}).

\bibitem{Sangouard2012}
\bibinfo{author}{Sangouard, N.} \& \bibinfo{author}{Zbinden, H.}
\newblock \bibinfo{title}{What are single photons good for?}
\newblock \emph{\bibinfo{journal}{Journal of Modern Optics}}
  \textbf{\bibinfo{volume}{59}}, \bibinfo{pages}{1458} (\bibinfo{year}{2012}).

\bibitem{Schlehahn2015}
\bibinfo{author}{Schlehahn, A.} \emph{et~al.}
\newblock \bibinfo{title}{Operating single quantum emitters with a compact
  {Stirling} cryocooler}.
\newblock \emph{\bibinfo{journal}{Review of Scientific Instruments}}
  \textbf{\bibinfo{volume}{86}}, \bibinfo{pages}{013113}
  (\bibinfo{year}{2015}).

\bibitem{Kuhlmann2013a}
\bibinfo{author}{Kuhlmann, A.~V.} \emph{et~al.}
\newblock \bibinfo{title}{Charge noise and spin noise in a semiconductor
  quantum device}.
\newblock \emph{\bibinfo{journal}{Nat Phys}} \textbf{\bibinfo{volume}{9}},
  \bibinfo{pages}{570} (\bibinfo{year}{2013}).

\bibitem{Gerard1998}
\bibinfo{author}{G\'erard, J.~M.} \emph{et~al.}
\newblock \bibinfo{title}{Enhanced spontaneous emission by quantum boxes in a
  monolithic optical microcavity}.
\newblock \emph{\bibinfo{journal}{Phys. Rev. Lett.}}
  \textbf{\bibinfo{volume}{81}}, \bibinfo{pages}{1110} (\bibinfo{year}{1998}).

\bibitem{Bleuse2011}
\bibinfo{author}{Bleuse, J.} \emph{et~al.}
\newblock \bibinfo{title}{Inhibition, enhancement, and control of spontaneous
  emission in photonic nanowires}.
\newblock \emph{\bibinfo{journal}{Phys. Rev. Lett.}}
  \textbf{\bibinfo{volume}{106}}, \bibinfo{pages}{103601}
  (\bibinfo{year}{2011}).

\bibitem{LundHansen2008}
\bibinfo{author}{Lund-Hansen, T.} \emph{et~al.}
\newblock \bibinfo{title}{Experimental realization of highly efficient
  broadband coupling of single quantum dots to a photonic crystal waveguide}.
\newblock \emph{\bibinfo{journal}{Phys. Rev. Lett.}}
  \textbf{\bibinfo{volume}{101}}, \bibinfo{pages}{113903}
  (\bibinfo{year}{2008}).

\bibitem{Dousse2008}
\bibinfo{author}{Dousse, A.} \emph{et~al.}
\newblock \bibinfo{title}{Controlled light-matter coupling for a single quantum
  dot embedded in a pillar microcavity using far-field optical lithography}.
\newblock \emph{\bibinfo{journal}{Phys. Rev. Lett.}}
  \textbf{\bibinfo{volume}{101}}, \bibinfo{pages}{267404}
  (\bibinfo{year}{2008}).

\bibitem{Reimer2012}
\bibinfo{author}{Reimer, M.~E.} \emph{et~al.}
\newblock \bibinfo{title}{Bright single-photon sources in bottom-up tailored
  nanowires}.
\newblock \emph{\bibinfo{journal}{Nat Commun}} \textbf{\bibinfo{volume}{3}},
  \bibinfo{pages}{737} (\bibinfo{year}{2012}).

\bibitem{Claudon2013}
\bibinfo{author}{Claudon, J.}, \bibinfo{author}{Gregersen, N.},
  \bibinfo{author}{Lalanne, P.} \& \bibinfo{author}{G\'{e}rard, J.-M.}
\newblock \bibinfo{title}{Harnessing light with photonic nanowires:
  Fundamentals and applications to quantum optics}.
\newblock \emph{\bibinfo{journal}{ChemPhysChem}} \textbf{\bibinfo{volume}{14}},
  \bibinfo{pages}{2353} (\bibinfo{year}{2013}).

\bibitem{Gazzano2013}
\bibinfo{author}{Gazzano, O.} \emph{et~al.}
\newblock \bibinfo{title}{Bright solid-state sources of indistinguishable
  single photons}.
\newblock \emph{\bibinfo{journal}{Nat Commun}} \textbf{\bibinfo{volume}{4}},
  \bibinfo{pages}{1425} (\bibinfo{year}{2013}).

\bibitem{Munsch2013}
\bibinfo{author}{Munsch, M.} \emph{et~al.}
\newblock \bibinfo{title}{Dielectric {GaAs} antenna ensuring an efficient
  broadband coupling between an {InAs} quantum dot and a gaussian optical
  beam}.
\newblock \emph{\bibinfo{journal}{Phys. Rev. Lett.}}
  \textbf{\bibinfo{volume}{110}}, \bibinfo{pages}{177402}
  (\bibinfo{year}{2013}).

\bibitem{Xu2007c}
\bibinfo{author}{Xu, X.} \emph{et~al.}
\newblock \bibinfo{title}{``plug and play'' single-photon sources}.
\newblock \emph{\bibinfo{journal}{Applied Physics Letters}}
  \textbf{\bibinfo{volume}{90}}, \bibinfo{pages}{061103}
  (\bibinfo{year}{2007}).

\bibitem{Haupt2010}
\bibinfo{author}{Haupt, F.} \emph{et~al.}
\newblock \bibinfo{title}{Fiber-connectorized micropillar cavities}.
\newblock \emph{\bibinfo{journal}{Applied Physics Letters}}
  \textbf{\bibinfo{volume}{97}}, \bibinfo{pages}{131113}
  (\bibinfo{year}{2010}).

\bibitem{Tiecke2015}
\bibinfo{author}{Tiecke, T.~G.} \emph{et~al.}
\newblock \bibinfo{title}{Efficient fiber-optical interface for nanophotonic
  devices}.
\newblock \emph{\bibinfo{journal}{Optica}} \textbf{\bibinfo{volume}{2}},
  \bibinfo{pages}{70} (\bibinfo{year}{2015}).

\bibitem{Cuche2009}
\bibinfo{author}{Cuche, A.} \emph{et~al.}
\newblock \bibinfo{title}{Near-field optical microscopy with a
  nanodiamond-based single-photon tip}.
\newblock \emph{\bibinfo{journal}{Opt. Express}} \textbf{\bibinfo{volume}{17}},
  \bibinfo{pages}{19969} (\bibinfo{year}{2009}).

\bibitem{Cuche2010}
\bibinfo{author}{Cuche, A.}, \bibinfo{author}{Mollet, O.},
  \bibinfo{author}{Drezet, A.} \& \bibinfo{author}{Huant, S.}
\newblock \bibinfo{title}{``{Deterministic}'' quantum plasmonics}.
\newblock \emph{\bibinfo{journal}{Nano Letters}} \textbf{\bibinfo{volume}{10}},
  \bibinfo{pages}{4566} (\bibinfo{year}{2010}).

\bibitem{Vamivakas2011}
\bibinfo{author}{Vamivakas, A.~N.} \emph{et~al.}
\newblock \bibinfo{title}{Nanoscale optical electrometer}.
\newblock \emph{\bibinfo{journal}{Phys. Rev. Lett.}}
  \textbf{\bibinfo{volume}{107}}, \bibinfo{pages}{166802}
  (\bibinfo{year}{2011}).

\bibitem{Note1}
\bibinfo{note}{We performed up to 3 cycles with the same device without loss of
  signal}.

\bibitem{Munsch2009}
\bibinfo{author}{Munsch, M.} \emph{et~al.}
\newblock \bibinfo{title}{Continuous-wave versus time-resolved measurements of
  {Purcell} factors for quantum dots in semiconductor microcavities}.
\newblock \emph{\bibinfo{journal}{Phys. Rev. B}} \textbf{\bibinfo{volume}{80}},
  \bibinfo{pages}{115312} (\bibinfo{year}{2009}).

\bibitem{Note2}
\bibinfo{note}{\protect \textit {Manuscript in preparation}}.

\bibitem{Friedler2009}
\bibinfo{author}{Friedler, I.} \emph{et~al.}
\newblock \bibinfo{title}{Solid-state single photon sources: the nanowire
  antenna}.
\newblock \emph{\bibinfo{journal}{Opt. Express}} \textbf{\bibinfo{volume}{17}},
  \bibinfo{pages}{2095--2110} (\bibinfo{year}{2009}).

\bibitem{Tame2013}
\bibinfo{author}{Tame, M.~S.} \emph{et~al.}
\newblock \bibinfo{title}{Quantum plasmonics}.
\newblock \emph{\bibinfo{journal}{Nat Phys}} \textbf{\bibinfo{volume}{9}},
  \bibinfo{pages}{329--340} (\bibinfo{year}{2013}).

\bibitem{Akimov2007}
\bibinfo{author}{Akimov, A.~V.} \emph{et~al.}
\newblock \bibinfo{title}{Generation of single optical plasmons in metallic
  nanowires coupled to quantum dots}.
\newblock \emph{\bibinfo{journal}{Nature}} \textbf{\bibinfo{volume}{450}},
  \bibinfo{pages}{402--406} (\bibinfo{year}{2007}).

\bibitem{Bracher2014}
\bibinfo{author}{Bracher, G.} \emph{et~al.}
\newblock \bibinfo{title}{Imaging surface plasmon polaritons using proximal
  self-assembled {InGaAs} quantum dots}.
\newblock \emph{\bibinfo{journal}{Journal of Applied Physics}}
  \textbf{\bibinfo{volume}{116}}, \bibinfo{pages}{033101}
  (\bibinfo{year}{2014}).

\bibitem{Andersen2011}
\bibinfo{author}{Andersen, M.~L.}, \bibinfo{author}{Stobbe, S.},
  \bibinfo{author}{Sorensen, A.~S.} \& \bibinfo{author}{Lodahl, P.}
\newblock \bibinfo{title}{Strongly modified plasmon-matter interaction with
  mesoscopic quantum emitters}.
\newblock \emph{\bibinfo{journal}{Nat Phys}} \textbf{\bibinfo{volume}{7}},
  \bibinfo{pages}{215--218} (\bibinfo{year}{2011}).

\bibitem{Fang2010}
\bibinfo{author}{Fang, Y.} \emph{et~al.}
\newblock \bibinfo{title}{Branched silver nanowires as controllable plasmon
  routers}.
\newblock \emph{\bibinfo{journal}{Nano Letters}} \textbf{\bibinfo{volume}{10}},
  \bibinfo{pages}{1950--1954} (\bibinfo{year}{2010}).

\bibitem{Wei2011a}
\bibinfo{author}{Wei, H.}, \bibinfo{author}{Wang, Z.}, \bibinfo{author}{Tian,
  X.}, \bibinfo{author}{Kall, M.} \& \bibinfo{author}{Xu, H.}
\newblock \bibinfo{title}{Cascaded logic gates in nanophotonic plasmon
  networks}.
\newblock \emph{\bibinfo{journal}{Nat Commun}} \textbf{\bibinfo{volume}{2}},
  \bibinfo{pages}{387} (\bibinfo{year}{2011}).

\bibitem{Houel2012}
\bibinfo{author}{Houel, J.} \emph{et~al.}
\newblock \bibinfo{title}{Probing single-charge fluctuations at a
  $\mathrm{GaAs}/\mathrm{AlAs}$ interface using laser spectroscopy on a nearby
  $\mathrm{InGaAs}$ quantum dot}.
\newblock \emph{\bibinfo{journal}{Phys. Rev. Lett.}}
  \textbf{\bibinfo{volume}{108}}, \bibinfo{pages}{107401}
  (\bibinfo{year}{2012}).

\bibitem{Karrai1995}
\bibinfo{author}{Karrai, K.} \& \bibinfo{author}{Grober, R.}
\newblock \bibinfo{title}{Piezoelectric tip-sample distance control for near
  field optical microscopes}.
\newblock \emph{\bibinfo{journal}{Applied Physics Letters}}
  \textbf{\bibinfo{volume}{66}}, \bibinfo{pages}{1842} (\bibinfo{year}{1995}).

\bibitem{Degen2008}
\bibinfo{author}{Degen, C.~L.}
\newblock \bibinfo{title}{Scanning magnetic field microscope with a diamond
  single-spin sensor}.
\newblock \emph{\bibinfo{journal}{Applied Physics Letters}}
  \textbf{\bibinfo{volume}{92}}, \bibinfo{pages}{243111}
  (\bibinfo{year}{2008}).

\end{thebibliography}


\end{document}